\documentclass[a4]{article}
\usepackage{comment}
\usepackage[dvips]{graphicx}
\title{Planar electric trap for neutral particles}

\author{Makoto \textsc{Morinaga}$^{1}$\thanks{E-mail address:
morinaga@ils.uec.ac.jp} and Tetsuo \textsc{Kishimoto}$^{2}$\\
\\
$^{1}$Institute for Laser Science, University of Electro-Communications
and JST-CREST, Japan\\
$^{2}$The Center for Frontier Science and
Engineering, University of Electro-Communications, Japan
}

\begin{document}
\maketitle
\abstract{
A new geometry to trap neutral particles with an ac electric field
using a simple electrodes structure is described.
In this geometry, all electrodes are placed on a single chip plane,
while particles are levitated
above the chip. This provides an easy construction
of the trap and a good optical access to the trap.
}

particle storage, atom optics, laser cooling


\section{Introduction}
To investigate the properties of particles such as atoms and molecules,
it is desirable to hold them in an isolated environment.
For this purpose, various kinds of traps have been developed.
Among them, magneto-optical trap (MOT) and magnetic trap (MT)
are successfully used for a variety of applications
\cite{metcalf}.
However they require the particles to have
either a closed transition (MOT) or
a magnetic moment (MT), which restrict the range
of particles to be trapped.
Electric trap (i.e. a trap that uses electric field),
on the other hand, has the advantage that
it can trap almost any kind of neutral particles.
It is easily shown that it is impossible to trap neutral particles
using a static electric field\cite{ketterle},
and an ac electric trap was proposed to overcome this difficulty.
\cite{shimizu,morinaga,katori}
Disadvantage of such electric trap is its shallow trap depth
(potential depth for sodium atoms is 190$\mu$K for an electric
field of 1kV\,mm$^{-1}$),
and one needs to make a small trap to increase the curvature of the
potential and hold particles against the gravity.
However ac electric traps proposed so far had solid structures
and the fabrication of a small trap was rather difficult.
For such technical reasons, ac electric traps were not realized until recently.
\cite{kishimoto,rempe,meijer_rb,meijer_molecule}

In this paper, we propose a new geometry of an ac electric trap
in which all electrodes are placed on a single plane, while
particles are trapped above that plane.
Such geometry enables one to fabricate the trap structure
in mass with high precision, which opens the possibility of 
making micro-structured atom and molecule chips that benefit from
the advantages of long decoherence time
\cite{henkel}, low electric power
consumption, and scalability of the electric traps.

\section{Principle of the dynamical trapping}
\begin{figure}
\begin{center}
\includegraphics[height=45mm]{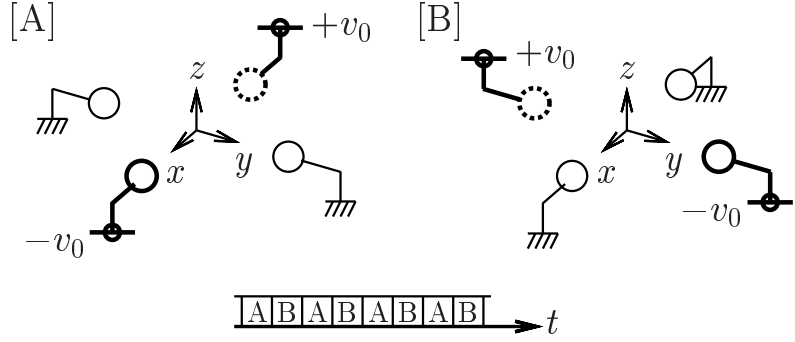}
\end{center}
\caption{Two pairs of electrodes are alternatively
switched on and off between phase A and B.
[phase A] Force is attractive in $yz$-plane and repulsive in $x$-direction.
[phase B] Attractive in $zx$-plane and repulsive in $y$-direction.
}
\label{principle}
\end{figure}

We start from the trap configuration described in [\cite{katori}],
schematically shown in fig. \ref{principle}, in which all electrodes
sit on a single plane ($xy$-plane).
Two pairs of electrodes are placed on $x$- and $y$-axis symmetrically
(i.e. at $(\pm s_0,0,0)$ and $(0,\pm s_0,0)$).
There are two phases in the operation of this trap:
In phase A (phase B) a voltage is applied on the pair of
electrodes on $x$-axis ($y$-axis).
The potential that a neutral particle feels under the
electric field ${\bf E(r)}$ is
\begin{equation}
V({\bf r})=-\frac 12\alpha|{\bf E(r)}|^2
\end{equation}
where $\alpha$ is the polarizability of the particle.
$\alpha$ is always positive for atoms or molecules in a stable state.
In phase A or B, the potential $V({\bf r})$ has the form
\begin{equation}
\begin{array}{llc}
V_A({\bf r})&=&\frac 12m\omega_0^2(-\eta_0 x^2+y^2+\xi_0 z^2)+V_0\\
V_B({\bf r})&=&\frac 12m\omega_0^2(+x^2-\eta_0 y^2+\xi_0 z^2)+V_0
\end{array}
\label{eqn0}
\end{equation}
up to 2nd order in ${\bf r}$.
Here, $m$ is the mass of the particle and $\omega_0$ is the
angular frequency of oscillation in $y$-direction in phase A
($x$-direction in phase B).
For spherical electrodes $\eta_0=2$ and $\xi_0=1$.\cite{inequality}
By switching between phase A and B alternatively
for time $T_A$ and $T_B$ (usually $T_A=T_B$),
particles are trapped around the origin $(0,0,0)$:
They are dynamically captured
in $xy$-plane (similar to the RF ion trap\cite{paul}), whereas
the confinement in $z$-direction is static.
Larger $\xi_0$ gives stronger confinement in $z$-direction,
while the stable region for the driving frequency $\Omega=2\pi/T$
is reduced with increasing $\eta_0$
($T\equiv T_A+T_B$ is the period of the applied voltage).
%
\section{How to lift up the trap point from the chip plane}
\begin{figure}
\begin{center}
\includegraphics[height=30mm]{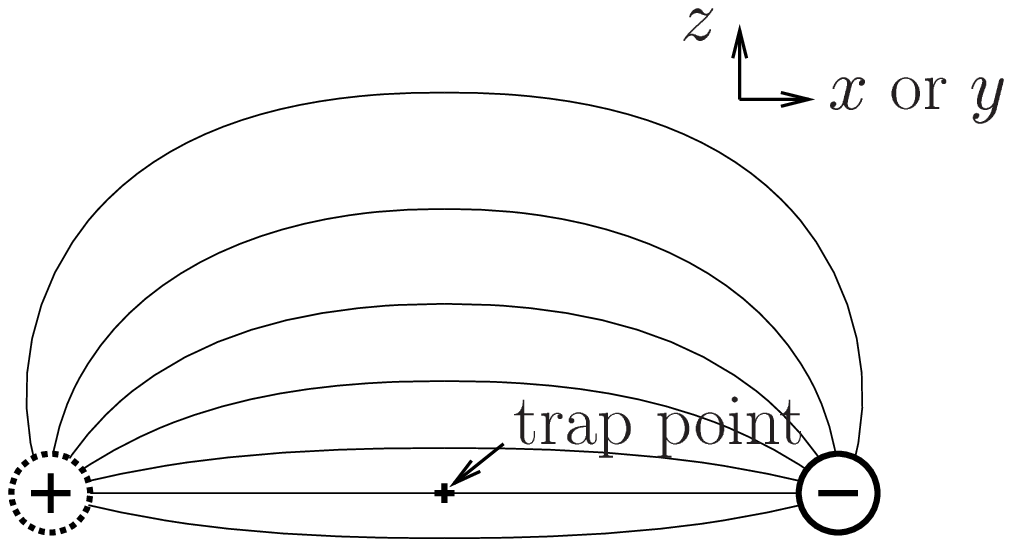}
\includegraphics[height=30mm]{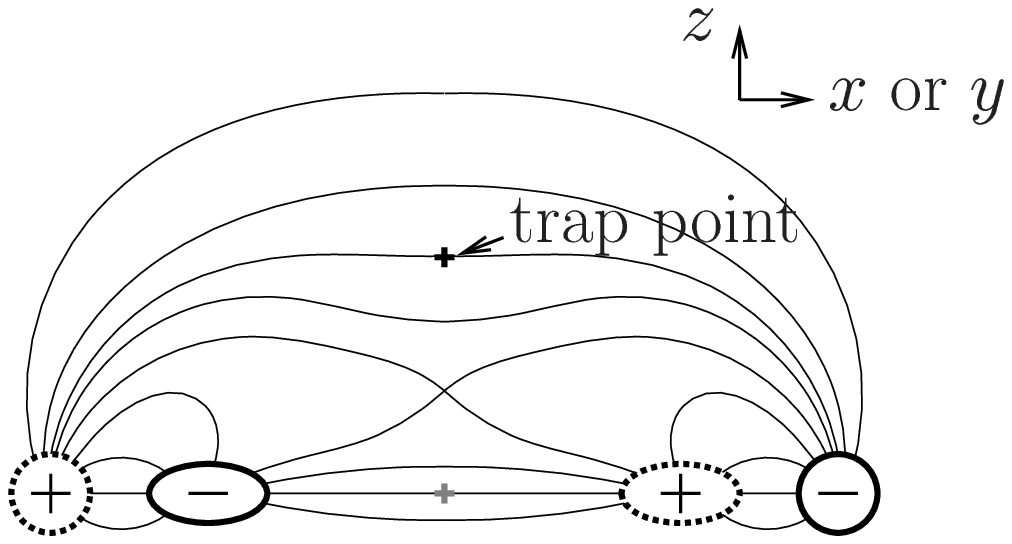}
\end{center}
  \caption{Trap point is the saddle point of the density of electric field lines.
By adding two additional charges, another saddle point is created above $xy$-plane
(right).
}
\label{lift}
\end{figure}

Particles are trapped at the saddle point of $|E({\bf r})|^2$.
In order to lift up this trapping point off the $xy$-plane,
we place additional electrodes in between the existing electrodes
to bend the field lines and make a saddle point other
than the origin (fig. \ref{lift}).
We still keep the electrodes configuration symmetric under
$x\leftrightarrow -x$,
$y\leftrightarrow -y$,
$x\leftrightarrow y$.
Now (\ref{eqn0}) becomes
\begin{equation}
\begin{array}{llc}
V_A({\bf r})&=&\frac 12m\omega^2\{-\eta x^2+y^2+\xi (z-h)^2\}+V,\\
V_B({\bf r})&=&\frac 12m\omega^2\{+x^2-\eta y^2+\xi (z-h)^2\}+V.
\end{array}
\label{eqn00}
\end{equation}
near the new saddle point $(0,0,h)$.
Let $\phi({\bf r})$ be the scalar potential.
In phase A (and similarily in phase B), because of
$\phi(x,y,z)=-\phi(-x,y,z)$,
$\phi(x,-y,z)=\phi(x,-y,z)$,
\begin{equation}
\left\{
 \begin{array}{l}
  E_x(x,y,z)=+E_x(-x,y,z)\\
  E_y(x,y,z)=-E_y(-x,y,z)\\
  E_z(x,y,z)=-E_z(-x,y,z)\\
  E_x(x,y,z)=+E_x(x,-y,z)\\
  E_y(x,y,z)=-E_y(x,-y,z)\\
  E_z(x,y,z)=+E_z(x,-y,z)
 \end{array}
\right.
\end{equation}
so that
$\partial_xE_x|_{(0,0,z)}=\partial_{y,z}E_y|_{(0,0,z)}=
\partial_{y,z}E_z|_{(0,0,z)}=
\partial_yE_x|_{(0,0,z)}=\partial_{x,z}E_y|_{(0,0,z)}=
\partial_yE_z|_{(0,0,z)}=0$.
And also $\partial_xE_z|_{(0,0,h)}=\partial_zE_x|_{(0,0,h)}=0$
from $\partial_z|{\bf E}|^2|_{(0,0,h)}=0$ and $\nabla\times{\bf E}=0$.
From these equalities, $\nabla^2|{\bf E}|^2\ |_{(0,0,h)}=2\sum_{i,j=\{x,y,z\}}
(\partial_iE_j|_{(0,0,h)})^2=0$,
and thus $\eta-\xi=1$ is still satisfied.

\subsection{Design procedure: step 1}
\begin{figure}
\begin{center}
\includegraphics[height=50mm]{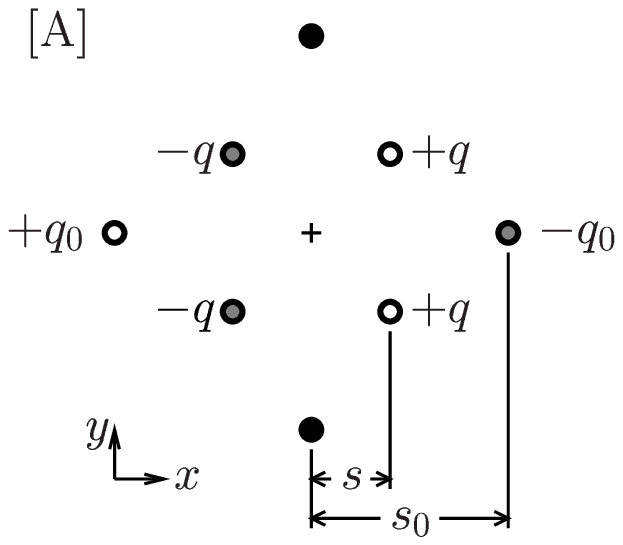}
\includegraphics[height=50mm]{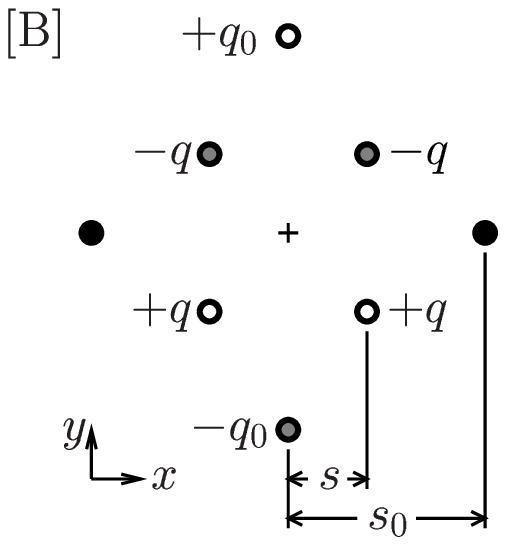}
\end{center}
  \caption{In addition to four outer point charges that corresponds
to the four electrodes of
of the original trap configuration, we place four inner point charges.
}
\label{d0}
\end{figure}

\begin{figure}
\begin{center}
(a)\includegraphics[width=70mm]{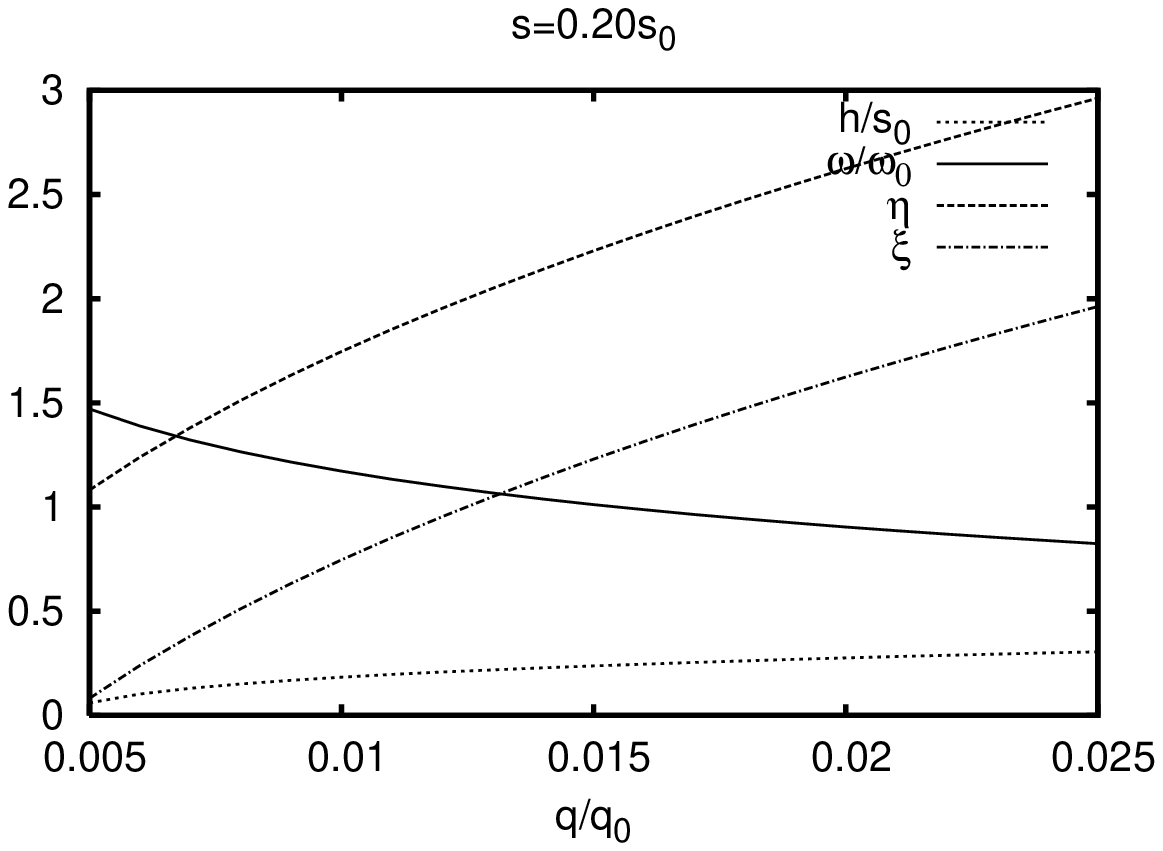}\\
(b)\includegraphics[width=70mm]{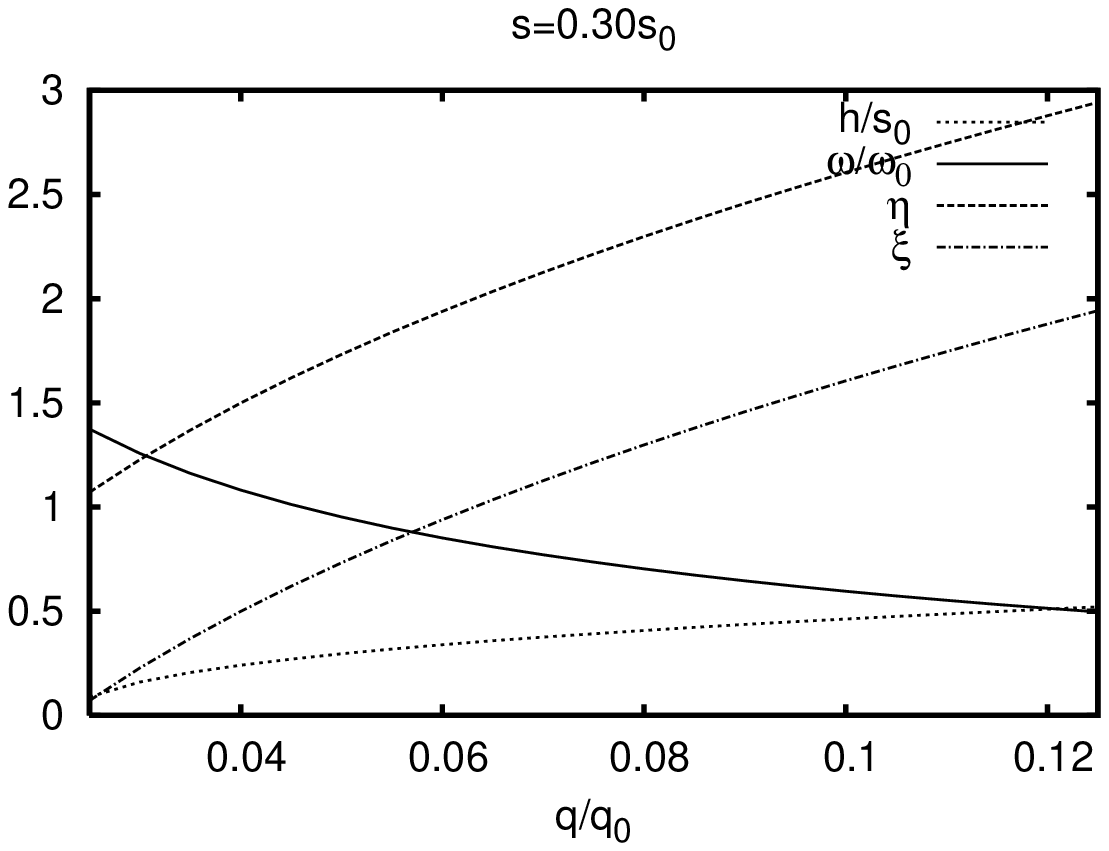}\\
(c)\includegraphics[width=70mm]{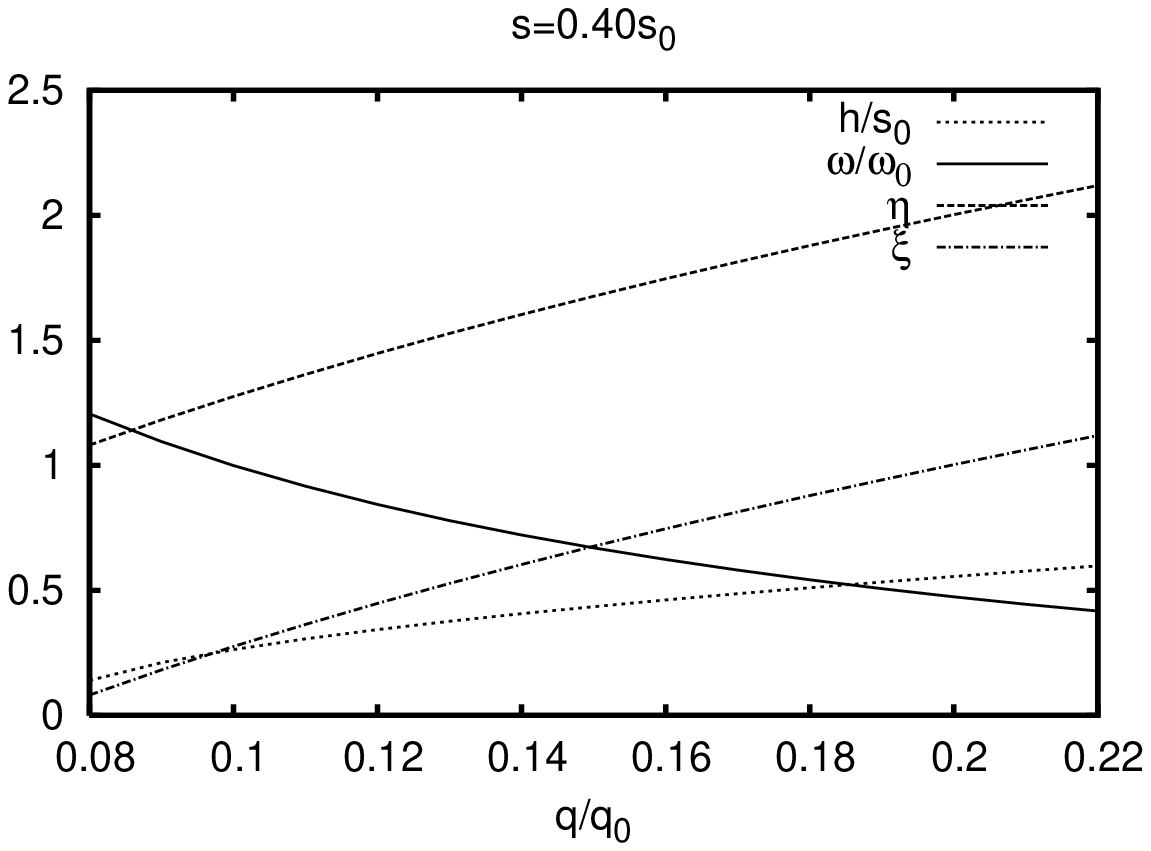}
\end{center}
\caption{Plot of the trap parameters: relative trap height
$\frac h{s_0}$, relative trapping strength
$\frac \omega{\omega_0}$, $\eta$, and $\xi$
are plotted as functions of $\frac q{q_0}$ for
$\frac s{s_0}=$0.2, 0.3, and 0.4.}
\label{params}
\end{figure}

\begin{figure}
\includegraphics[height=50mm]{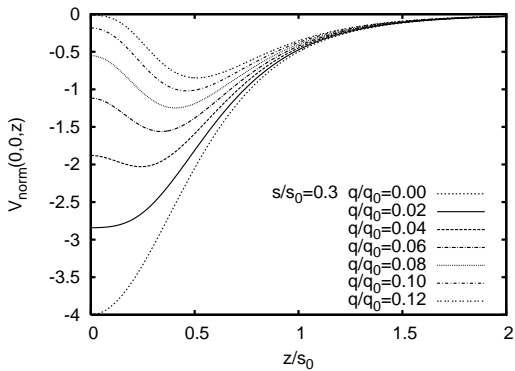}
  \caption{Plot of the Stark potential along z-axis for different $q$.}
\label{exampleq}
\end{figure}

\begin{figure}
\includegraphics[height=50mm]{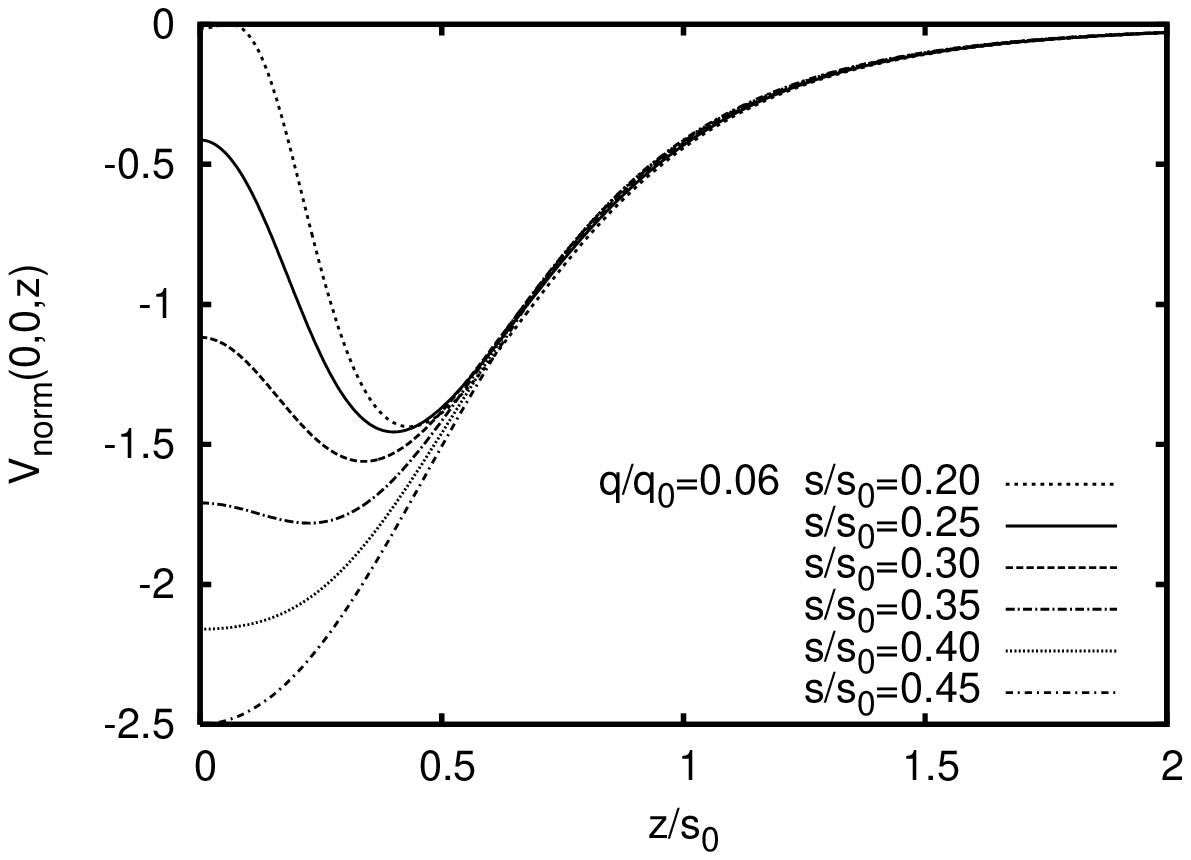}
  \caption{Plot of the Stark potential along z-axis for different $s$.}
\label{examples}
\end{figure}

\begin{figure}
\includegraphics[height=50mm]{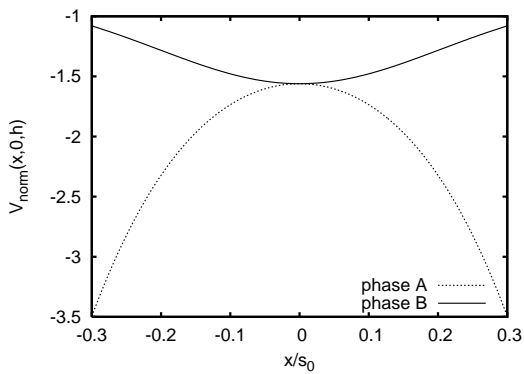}
  \caption{Plot of the Stark potential along x-direction
for phase A and phase B ($s=0.3s_0$, $q=0.06q_0$).}
\label{examplex}
\end{figure}

First, we consider electrodes as point charges and place them
on $xy$-plane as shown in
fig. \ref{d0}.
Four outer point charges placed at
${\bf s_{out}}=(\pm s_0,0),\ (0,\pm s_0)$
correspond to the four electrodes
of the original configuration (fig.\ref{principle}) and
we put charges of $\pm q_0$ or 0 at appropriate phase.
Four inner point charges at
${\bf s_{in}}=(\pm s,\pm s),\ (\pm s,\mp s)$
are added to the original configuration with charges
$\pm q$ depending on the phase (see figure).
In fig. \ref{params}
we plot
parameters $\frac h{s_0}$, $\frac\omega{\omega_0}$, $\xi$
that appear in (\ref{eqn00})
as functions of $\frac q{q_0}$ for several $\frac s{s_0}$ values
($\eta$ is calculated from $\xi$ as $\eta=\xi+1$).
Here $\omega_0$ is $\omega$ for the original configuration,
i.e. $\omega_0=\sqrt{\left.\frac 1m\partial_y^2V_A\right|_{{\bf r}=0,\,q=0}}
=\frac 2{s_0^2}\sqrt{\frac{3\alpha}m}
\frac{q_0}{4\pi\epsilon_0s_0}$.
Normalized potential $V_{norm}({\bf r})\equiv\frac {2s_0^2}\alpha
\left(\frac{4\pi\epsilon_0s_0}{q_0}\right)^2
V({\bf r})$ is plotted along $z$-axis for different $q$
in fig. \ref{exampleq}, for different $s$ in fig. \ref{examples},
and along $x$- ($y$-) direction in fig. \ref{examplex}.
We choose $\frac s{s_0}=0.3$ and $\frac q{q_0}=0.06$ to proceed the design
procedure further.
Relevant parameters for $\frac s{s_0}=0.3$ and $\frac q{q_0}=0.06$
are compared with those
for the original configuration in
table \ref{compare}
({\it see section \ref{polar} for $\chi$}).
\begin{table}
\begin{center}
\begin{tabular}{|l||l|l|l|l|l|}
\hline
$\frac q{q_0}$ & $\frac h{s_0}$ & $\frac \omega{\omega_0}$
 & $\eta$ & $\xi$ & $\chi$\\
\hline
\hline
0.06 & 0.34 & 0.85 & 1.94 & 0.94 & 0.64\\
\hline
\hline
(0.0) & (0.0) & (1.0) & (2.0) & (1.0) & (1.0)\\
\hline
\end{tabular}
\end{center}
\caption{parameter comparison table for $s=0.3s_0$.}
\label{compare}
\end{table}
\subsection{Design procedure: step 2}
\begin{figure}
(a)\includegraphics[height=55mm]{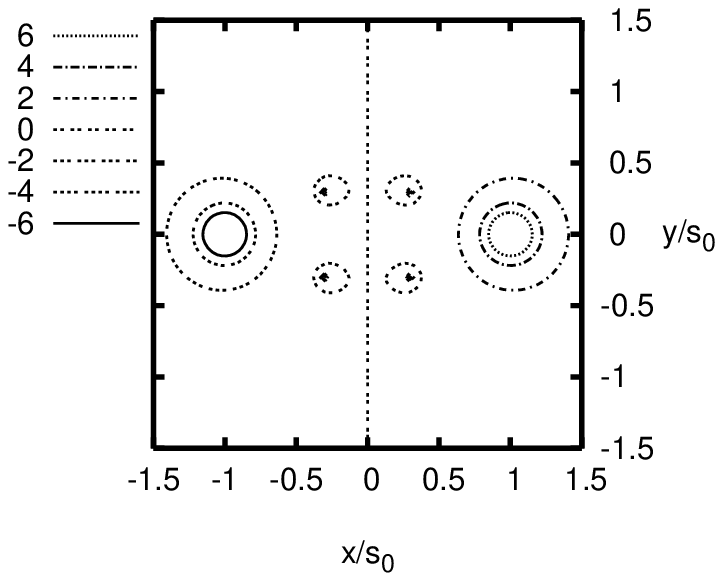}
(b)\includegraphics[height=55mm]{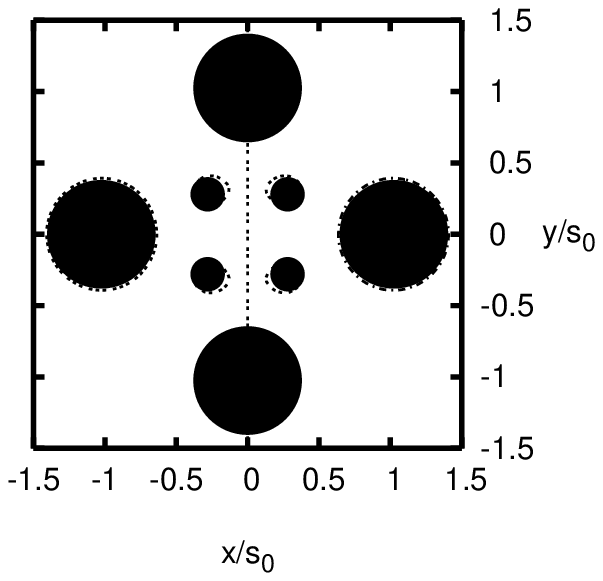}
  \caption{ (a) Cross section of the equipotential surface in phase A:
normalized scalar potential $\phi_{norm}({\bf r})$
for $s=0.3s_0$ and $q=0.06q_0$
at the chip surface ($z=0$) is plotted as a function of $x$ and $y$.
(b) Disc shaped electrodes shown in black imitate
the equipotential surfaces.
Inner electrodes are connected to the ground.
}
\label{equipotential}
\end{figure}

Now we are going to replace the point charges by electrodes
of finite size.
For that, first we calculate the equipotential surface of the electric field.
Figure \ref{equipotential}a is the plot
of the normalized scalar
potential $\phi_{norm}({\bf r})\equiv
\frac{4\pi\epsilon_0s_0}{q_0}\phi({\bf r})$ in the chip plane ($z=0$).
Equipotential surfaces of $\phi_{norm}=\pm a\ \ (a>0)$
 around the outer
point charges at ${\bf s_{out}}$
are nearly spherical for large $a$ and small $q/q_0$,
and can be well imitated
by a disc shaped electrodes of radius $r_{out}=\frac {2s_0}{2a+1}$
centered at ${\bf s_{out}}$.
However smaller $a$
gives higher electric field assuming that the voltage applied
on the electrodes is fixed.
In fig. \ref{equipotential}b we choose $a=2.0$
($r_{out}=0.4s_0$).
We apply to these outer electrodes voltage of $\pm v_0$ or 0 
at appropriate phase: 
$\frac{4\pi\epsilon_0s_0}{q_0}=\frac a{v_0}$
(i.e. $V({\bf r})=\frac\alpha{2a^2}\left(\frac{v_0}{s_0}\right)^2
V_{norm}({\bf r})$).
Equipotential surfaces of $\phi_{norm}=0$ around the inner
point charges at ${\bf s_{in}}$
are again nearly spherical for $\frac q{q_0}\ll \frac s{s_0}$,
and the inner point charges can be
replaced by disc shaped electrodes of radius
$r_{in}=\frac q{q_0}\{(s_0^2-2s_0s+2s^2)^{-\frac 12}
-(s_0^2+2s_0s+2s^2)^{-\frac 12}\}^{-1}$
($r_{in}=0.11s_0$ for $q=0.06q_0$, $s=0.3s_0$)
centered at ${\bf s_{in}}$
which are always connected to the ground (0 voltage).
%

\section{Trapping particles under the gravity of the earth}
In this section, we give some practical values to trap neutral atoms
and dielectric spheres on the earth.
We orient $z$-axis along the gravity direction.
We set a condition that the gravitational sag should be smaller
than $\sigma s_0$ (with $\sigma\ll 1$):
$\frac g{\xi\omega^2}\le\sigma s_0$.
This gives $\omega_0\ge
\sqrt{\frac g{\sigma s_0\xi\frac \omega{\omega_0}}}$.
On the other hand $\omega_0=\frac 2{s_0^2}
\sqrt{\frac{3\alpha}m}\frac{v_0}a$.
We fix here the trap design as $\frac s{s_0}=0.3$,
$\frac q{q_0}=0.06$ and $a=2.0$ which gives $\frac \omega{\omega_0}=0.85$,
$\xi=0.94$. We also put $\sigma=0.1$ and $g=9.81\,m\,s^{-2}$.

First we discuss on trapping neutral atoms. We choose
sodium atoms as an example:
%
$\alpha=2.68\cdot 10^{-39}\mathrm{F\,m^2}$,
\cite{pritchard}
$m=3.82\times 10^{-26}$kg, and thus
$\frac\alpha m=7.0\times 10^{-14}\mathrm{F\,m^2kg^{-1}}$.
If we set $s_0=0.5$mm, then the condition on the gravitational
sag requires $\omega_0\ge 2\pi\times 79$Hz.
This will be satisfied with the applied voltage of $v_0\ge 270$V.
Setting $v_0=270$V, then $\omega=2\pi\times 67$Hz and,
using fig. \ref{stable_region},
the trap should be stable for
$7.8\mathrm{ms}<T<9.5$ms with $T_A=T_B=\frac T2$.

Next we discuss on trapping of a dielectric material.
The polarizability of a dielectric sphere of radius $a$
and dielectric constant $\epsilon$ is 
$\alpha=4\pi\epsilon_0
\frac{\epsilon-\epsilon_0}{\epsilon+2\epsilon_0}a^3$
(mass is $m=\frac{4\pi}{3}\rho\,a^3$ where $\rho$ is the density)
so that $\frac\alpha m=
\frac{3\epsilon_0}\rho
\frac{\epsilon-\epsilon_0}{\epsilon+2\epsilon_0}$
is independent of its size.
Polystyrene($\epsilon=2.5\epsilon_0$,
$\rho=1.05\times 10^{3}\mathrm{kg\,m^{-3}}$), for example,
has the value
$\frac\alpha m=8.4\times 10^{-15}\mathrm{F m^2kg^{-1}}$,
which is roughly one order smaller than that of
sodium atoms.

\section{Trapping of polar particles}
\label{polar}
\begin{figure}
\begin{center}
\includegraphics[width=90mm]{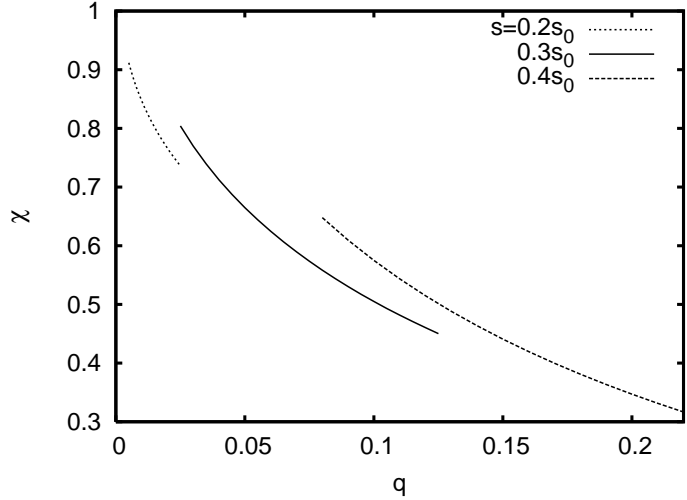}
\end{center}
\caption{
Plot of $\chi$ as a function of $\frac q{q_0}$ for several
$\frac s{s_0}$ values.
}
\label{chi}
\end{figure}

In the case of polar particles of permanent dipole
moment ${\bf\mu}$, 
the force under the electric field can be written as,
assuming the direction of the dipole moment is always oriented
parallel to the electric field ({\it see appendix \ref{adiabatic}}),
\begin{equation}
\begin{array}{rl}
F_i({\bf r})=&\sum_j\mu_j\partial_jE_i({\bf r})\\
=&\sum_j\mu\frac{E_j({\bf r})}{E({\bf r})}\partial_jE_i({\bf r})\\
=&\mu\sum_j\frac{E_j({\bf r})}{E({\bf r})}\partial_iE_j({\bf r})\\
=&\frac{\mu}{2E({\bf r})}\partial_i\{E({\bf r})^2\}\\
=&\mu\partial_iE({\bf r})
\end{array}
\end{equation}
Thus the force is derived from the potential
\begin{equation}
V({\bf r})=-\mu E({\bf r})
\end{equation}
Near the trap center ${\bf r_c}$,
$E({\bf r})=\frac{E({\bf r})^2+E({\bf r_c})^2}{2E({\bf r_c})}$, so that
the dynamics of the polar particles can be argued in the same
way as that of the nonpolar particles
by replacing $\alpha$ with $\frac\mu{E({\bf r_c})}$.
We define a dimensionless parameter
$\chi\equiv\frac{E({\bf r_c})}{E_0}$ where
$E_0\equiv E(0)|_{q=0}=\frac{q_0}{4\pi\epsilon_0s_0}
|\nabla\phi_{norm}(0)|_{q=0}=\frac{2v_0}{as_0}$.
Now $\omega_0=\frac 2{s_0^2}
\sqrt{\frac{3}m\frac\mu{\chi E_0}}\frac{v_0}a
=\sqrt{\frac{6\mu v_0}{\chi a m s_0^3}}$.
A plot of $\chi$ is given in fig.\ref{chi}.
For a particle of bulk material of permanent polarization
$P$ and density $\rho$, $\frac\mu m=\frac P\rho$ so that
$\omega_0
=\sqrt{\frac{6 P v_0}{\chi a \rho s_0^3}}$ is independent
of its shape and size.
To give practical parameters, we again use the trap design
$s=0.3s_0$, $q=0.06q_0$, and $a=2.0$,
and consider trapping of $\mathrm{BaTiO_3}$ micro particles
($P=0.26\mathrm{Cm^{-2}}$ and $\rho=5.5\times 10^3
\mathrm{kg\,m^{-3}}$).
If we set $s_0=3.0$mm, then the condition on the gravitational sag
with $\sigma=0.1$ requires $\omega_0\ge 2\pi\times 32$Hz which
implies $v_0\ge 5.0$V. At $v_0=5.0$V, $\omega=2\pi\times 27$Hz and,
from fig. \ref{stable_region}, the
trap should be stable for $19\mathrm{ms}<T<23$ms.
\section{Conclusions and outlook}
\begin{figure}
\includegraphics[width=80mm]{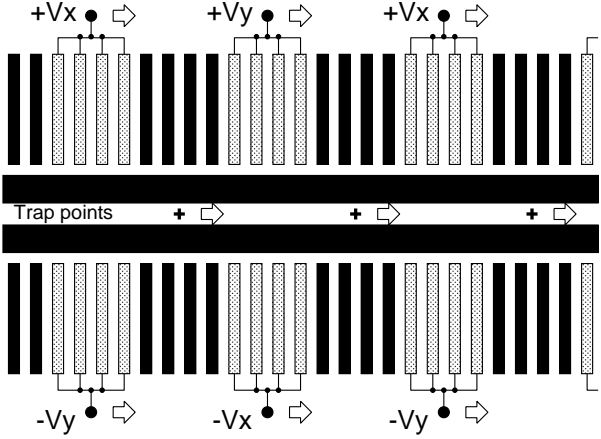}
\caption{
Possible design of a particle conveyor.
'+' marks show the trapping points levitated off the chip surface.
Electrodes shown in black are connected to the ground.}
\label{conveyor}
\end{figure}

We have presented a new design of ac electric trap
for neutral particles and have shown
that it is possible to make a trapping point above the chip surface with
planar electrodes structure.
We have also discussed about the feasibility of this design by showing
typical parameters for trapping neutral atoms, dielectric spheres,
and polar particles.
The parameters for this trap were comparable with those demonstrated in
[\cite{kishimoto}].
In the final step of the design, we have used disc shaped electrodes to
imitate the equipotential surface of the electric field.
\cite{dielectric}
To analyse its consequence, we are now calculating
the electric field using numerical methods.
The result is still preliminary, but it has turned out that the
field near the trapping point is not so much dependent on the
shape of the inner electrodes.
Thus, as an application of our trap, a particle conveyor
such as shown in fig. \ref{conveyor} should be also possible.

\vspace{1cm}
{\bf\large acknowledgement}\\
This work was partly supported by the 21st Century COE
program of the University of Electro-Communications on
"Coherent Optical Science" supported by the Ministry of
Education, Culture, Sports, Science and Technology.

\appendix
\section{Stability}
In this section, we investigate the stable condition of
the dynamical confinement.
The equations of motion are separated in $x$-, $y$-,
$z$-directions
so that we explore the motion in $x$-direction only.
We follow the procedure described in [\cite{morinaga}]
and define a state vector $X(t)$ as
\begin{equation}
 X(t)=\left(
\begin{array}{c}
 x(t)\\v(t)
\end{array}
\right)
\end{equation}
where $x(t)$ ($v(t)$) is the position (velocity) of
the particle at time t.
Time evolution of $X$ is described by a time evolution
matrix $U(t)$: $X(t)=U(t)X(0)$.
\begin{equation}
 U(t)=\left(
\begin{array}{cc}
 x^{(1)}(t) & x^{(2)}(t)\\
 v^{(1)}(t) & v^{(2)}(t)
\end{array}
\right)
\end{equation}
with $U(0)=I$ ($I$ is the unit matrix).
\begin{figure}
\begin{center}
\includegraphics[width=80mm]{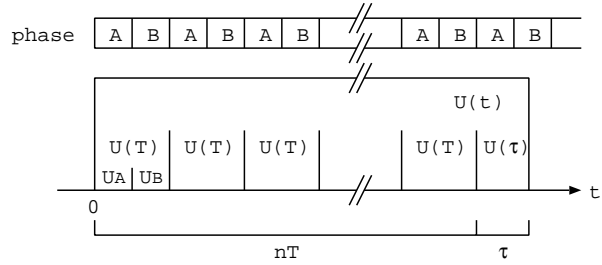}
\end{center}
\caption{
Time evolution matrix $U$.
}
\label{evolution}
\end{figure}
We assume that the phase A starts at $t=0$ (fig. \ref{evolution})
so that $U(nT+\tau)=U(\tau) U(T)^n$ with
$U(T)=U_BU_A$ where $U_A$ ($U_B$) is the time evolution matrix for
the phase $A$ ($B$).
\begin{equation}
 U_A=\left(
\begin{array}{cc}
 \cosh\sqrt\eta\omega T_A &
   \frac 1{\sqrt\eta\omega}\sinh\sqrt\eta\omega T_A\\
 \sqrt\eta\omega\sinh\sqrt\eta\omega T_A &
   \cosh\sqrt\eta\omega T_A
\end{array}
\right)
\end{equation}
\begin{equation}
 U_B=\left(
\begin{array}{cc}
 \cos\omega T_B & \frac 1{\omega}\sin\omega T_B\\
 -\omega\sin\omega T_B & \cos\omega T_B
\end{array}
\right)
\end{equation}
with $T_A=T_B=\frac T2$.
Let $\lambda$ be the eigenvalue of $U(T)$.
\begin{equation}
 \lambda^2-2\beta\lambda+1=0
\label{eigenvalue}
\end{equation}
with $\beta=\frac 12\mathrm{Tr}\ U(T)
=\cosh\sqrt\eta\omega T_A\cos\omega T_B+
\frac {\eta-1}{2\sqrt\eta}
\sinh\sqrt\eta\omega T_A\sin\omega T_B
$.
From (\ref{eigenvalue})
\begin{equation}
 |\lambda|=\left\{
  \begin{array}{cl}
   1\ \ &(|\beta|\le 1)\\
   |\beta|\pm\sqrt{\beta^2-1}&(|\beta|> 1)\\
  \end{array}
\right.
\end{equation}
The trap is stable if and only if $|\beta|\le 1$.
\begin{figure}
\begin{center}
\includegraphics[width=100mm]{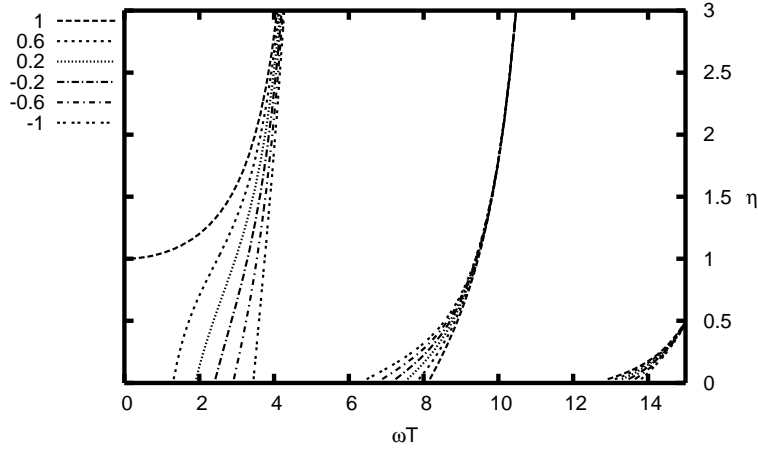}
\end{center}
\caption{
Plot of $\beta$ as a function of $\omega T$ and $\eta$.
The trap is stable in the region where $|\beta|<1$.}
\label{stable_region}
\end{figure}
In fig. \ref{stable_region} we plot $\beta$ as a function
of $\omega T$ and $\eta$ ($T_A=T_B=\frac T2$)
to show the stable region.
\subsection{Unharmonicity}
So far we have not included the unharmonic terms in the potential.
In the original Stark chip configuration, the lowest unharmonic term was
4th order because of the $z\rightarrow -z$ symmetry. In our configuration,
due to the absence of this symmetry, 3rd order terms are present.
However these terms have the form
\begin{equation}
 V_3(x,y,z)=(\gamma_1\,x^2+\gamma_2\,y^2)(z-h)+\gamma_3\,(z-h)^3
\end{equation}
and are still harmonic in the plane $z=const.$
(i.e. the plane in which
the confinement is dynamical), we expect that the 3rd order terms
do not reduce the stable region in phase space drastically.
%
%
%
\section{Trap tightness}
\begin{figure}
\begin{center}
\includegraphics[width=65mm]{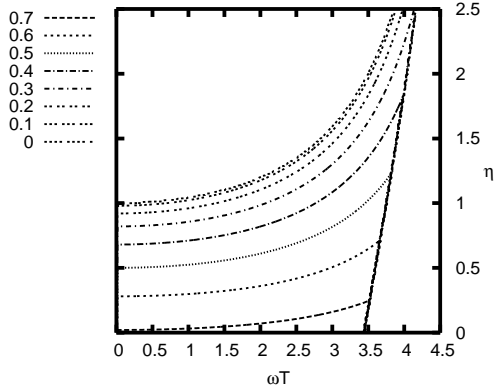}
\end{center}
\caption{
Plot of $\frac{\omega_{eff}}{\omega}$ as a function of
$\omega T$ and $\eta$ ($T_A=T_B=\frac T2$).
}
\label{omega_eff}
\end{figure}

To estimate the tightness of the dynamical confinement,
we apply a constant force $F_0$ in $x$-direction
on the particle in the trap and calculate the shift of the
trap center (i.e. the time average of the position of the
particle).
Let $W(t)$ be the state vector at time $t$ with the initial condition
$W(0)=0$.
Then the time evolution of a general state vector is now
$X(t)=U(t)X(0)+W(t)$.
\begin{equation}
\begin{array}{ll}
W(2T)&=U(T)W(T)+W(T)\\
W(3T)&=U(T)\{U(T)W(T)+W(T)\}+W(T)\\
W(nT)&=\{U(T)^{n-1}+U(T)^{n-2}+...+U(T)+1\}W(T)\\
&=\frac{1-U(T)^n}{1-U(T)}W(T)
\end{array} 
\end{equation}
\begin{equation}
W(nT+\tau)
=U(\tau)\frac{1-U(T)^n}{1-U(T)}W(T)+W(\tau)
\end{equation}
\begin{equation}
X(nT+\tau)
=U(\tau)U(T)^nX(0)+U(\tau)\frac{1-U(T)^n}{1-U(T)}W(T)+W(\tau)
\end{equation}
To evaluate the time average of $X(t)$, we put $t=nT+\tau$
with $n=0,1,2,...$ and $0\le\tau<T$, and
take the average over $n$ and $\tau$.
Because $\overline{U(T)^n}=0$ inside the stable region, we obtain
\begin{equation}
\overline{X(nT+\tau)}
=\left(
 \begin{array}{c}
  \overline{x(nT+\tau)} \\ \overline{v(nT+\tau)}
 \end{array}
\right)
=\overline{U(\tau)}\frac 1{1-U(T)}W(T)+\overline{W(\tau)}
\label{t_average}
\end{equation}
After a lengthy calculation, (\ref{t_average}) yields
\begin{equation}
\begin{array}{c}
 \overline{x(nT+\tau)}
=x_0
\left\{
\frac{-T_A+\eta T_B}{\eta T}
+\frac{(\cosh\sqrt\eta\omega T_A-1)\sin\omega T_B
-\sqrt\eta(\cos\omega T_B-1)\sinh\sqrt\eta\omega T_A}
{(1-\beta)\omega T}
\left(\frac 1\eta+1\right)^2
\right\}
\end{array}
\label{t_average2}
\end{equation}
and  $\overline{v(nT+\tau)}=0$,
where $x_0=\frac{F_0}{m\omega^2}$.
In the limit $\omega T\rightarrow 0$, $\overline{x(nT+\tau)}
=\frac 2{1-\eta}\,x_0$ as expected.
The effective trap frequency is calculated as
$\omega_{eff}=\sqrt{\frac{x_0}{\overline{x(nT+\tau)}}}\ \omega$.
In fig. \ref{omega_eff} we plot $\frac{\omega_{eff}}{\omega}$
for $T_A=T_B=\frac T2$ as a function of
$\omega T$ and $\eta$.
From this, we see that $\frac{\omega_{eff}}{\omega}=0.25\sim 0.33$
is obtained for $\eta=1.5\sim 2.0$.
\section{Dynamics of the polarization orientation of a polar particle}
\label{adiabatic}
Consider a polar particle of size $L$ and volume $u$ ($u\sim L^3$)
that consists from a bulk material under the electric field $E$.
Let $\theta$ be the angle between the electric field $E$ and the
polarization $P$ of the material. Then
\begin{equation}
 I\,\ddot\theta=N=-uPE\,\sin\theta
\end{equation}
where $I$ is the moment of inertia and $N$ is the torque.
$I\sim\rho L^5$ where $\rho$ is density of the material
($I=\frac\pi{60}\rho L^5$ for a sphere of diameter $L$).
Assuming $\theta$ is small,
$\ddot\theta\sim-\frac{PE}{\rho\,L^2}\theta$
(sphere: $\ddot\theta=-10\frac{PE}{\rho\,L^2}\theta$).
Angular frequency of oscillation of the polarization
direction is thus $\omega_{osc}\sim\sqrt{\frac{PE}\rho}\frac 1L$
(sphere: $\omega_{osc}=\sqrt{\frac{10PE}\rho}\frac 1L$).
If the switching time between phase A and B is longer than
$\omega_{osc}^{-1}$, then the direction of the polarization
adiabatically follows that of the electric field.
For $\mathrm{BaTiO_3}$
({\it see section \ref{polar}})
 with $L=1\mu$m and $E=1\mathrm{V\,mm^{-1}}$,
$\omega_{osc}\sim 2\times 10^5\mathrm{s^{-1}}$
(sphere: $\omega_{osc}=7.8\times 10^5\mathrm{s^{-1}}$).
%

%
%
%


\begin{thebibliography}{99}
\bibitem{dielectric}
The presence of the chip substrate on which electrodes
are placed does not affect the electric field in the free space side
as far as electrodes are thin and the substrate is thick compared
to the electrodes structure.
\end{thebibliography}
\end{document}